\documentclass[twocolumn,showpacs,preprintnumbers,prd,nofootinbib]{revtex4-1}
\usepackage{latexsym}
\usepackage{graphicx}
\usepackage{amsmath}
\usepackage{amsfonts}
\usepackage{amssymb}

\begin{document}
\title{Study of symmetry breaking of charged scalar field: Hydrodynamic version}



\author{T Matos$^1$ and M A Rodr\'iguez-Meza$^2$}

\affiliation{$^1$Departamento de F\'\i sica, Centro de Investigaci\'on 
y de Estudios Avanzados del IPN, 
A.P. 14-740, 07000 Mexico City, 
M\'{e}xico}

\affiliation{$^2$Departamento de F\'{\i}sica, Instituto Nacional de
Investigaciones Nucleares, 
Apdo. Postal 18-1027, M\'{e}xico D.F. 11801,
M\'{e}xico. E-mail: marioalberto.rodriguez@inin.gob.mx
}





\begin{abstract}
We rewrite the Klein-Gordon (KG) equation for a complex scalar field as a new Gross-Pitaevskii (GP)-like equation. 
The potential of the scalar field is a mexican-hat potential and the field is in a thermal bath with 
one loop contribution. 
We interpret the new GP equation as a finite temperature generalization of the GP equation for a charged field. 
We find its hydrodynamic version as well and using it, we derive the corresponding thermodynamics. 
We also obtain a generalized first law for a charged Bose-Einstein Condensate (BEC). 
\end{abstract}

\maketitle

\section{Introduction}\label{sec:Intro}
Now a days, several astronomical observations give evidence for the existence of the so called dark matter, occupying approximately the 22\% of the universe, the rest is contained in another dark component called the dark energy with about 68\%, usually associated to a fundamental cosmological constant and the baryonic matter with less than 5\%. 
The standard model to study the universe and its evolution is a model based on assuming that dark matter is a cold component with an equation of state of dust type, and a cosmological constant ($\Lambda$) for the other dark component (see for example \cite{LCDM} for a review). 
This model is usually referred as $\Lambda$CDM and is able to feet almost all the observations of the universe we have till now at cosmological scales with a very good accuracy. However, this model has some issues at small scales, like, among others, its prediction of cuspy halos of galaxies. This model predicts that the dark matter at the center of the galaxies must have a cusp density profile, while observations tell us that galaxies prefer to have almost constant density profiles in the center, specially dwarf or low surface brightness galaxies.
Alternativelly, a scalar field has been proposed as a dark matter model (see \cite{SFDM01, SFDM02} or \cite{Suarez2013} for a review). In this model the nature of dark matter is a fundamental scalar field, where the predictions at cosmological scales coincide with the ones of the LCDM model, but it predicts a flat central density profile as observed in galaxies. In short, in the theoretical Lagrangian, we add two terms, a kinetic term and a potential term for the scalar field. We then fix the form of the potential and with the resultant equations we try to explain some observed astronomical data, like rotation curves of spiral galaxies. In particular, this scalar field model predicts that the galaxy dark halo is core-type. We refer the reader to  \cite{Suarez2013}  to see more details and comparisons with observations and 
with the $\Lambda$CDM model.

In this work we generalized the model by considering that the scalar field is a complex field and also we consider an electromagnetic term in the Lagrangian. With this model we derive the hydrodynamical version where the potential exhibits  a symmetry breaking (SB) and a possible thermodynamics (see also \cite{Matos2011} for the real case). Symmetry breaking is normally associated to phase transitions in other areas of physics, therefore its importance, in this work we would like to see whether the symmetry breakdown of the original Lagrangian symmetry has some effect in the structure formation formation of the universe.

We organise
our work in the following form: In the next section we present a general Lagrangian for complex scalar field which includes a term with electromagnetic fields and show how the symmetry appears and how can be broken. 
Next we show how the Klein-Gordon equation transforms in a generalized Gross-Pitaevskii equation. Three sections follow where we analyze the hydrodynamical version of the generalized Gross-Pitaevskii equation, its Newtonian limit and the thermodynamics of the system.
In the final section we write our conclusions of this work.

\section{General Lagrangian of the charged scalar field and the gauge symmetry breaking}\label{sec:SB}

The general Lagrangian model for a charged scalar field (SF) having a local U(1) symmetry is, 
  \begin{equation}
   {\cal L}=\left(\nabla_{\mu}\Phi+\mathrm{i} eA_\mu\Phi\right)\left(\nabla^{\mu}\Phi^*-\mathrm{i} eA^\mu\Phi^*\right)
   +V(\Phi\Phi^*)-\frac{1}{4}F^{\mu\nu}F_{\mu\nu}
  \label{eq:L}
  \end{equation}
where $V$ is the scalar field potential which is a double-well interacting Mexican-hat potential for a complex SF $\Phi(\mathbf{x},t)$. The scalar field is in thermal equilibrium with a reservoir at temperature $T$. The thermal interaction 
between the scalar field and the thermal bath is computed 
up to one loop of correction, and is given by \cite{Matos2011,Kolb1990},
  \begin{equation}
   V(\Phi\Phi^*)=- m^2\Phi\Phi^*+\frac{\lambda}{2}(\Phi\Phi^*)^2+\frac{\lambda}{4}T^2\Phi\Phi^* -\frac{\pi^2}{90}T^4,
  \label{eq:V}
  \end{equation}
This result includes both quantum and thermal contributions. Parameter $m$ is the mass of the scalar field, $\lambda$ gives us the scalar field self interaction and will be related to a ``pressure'', in the sense that there is an equation that relates both $p \propto \rho$.
\begin{figure}
(a)
\includegraphics[width=1.5in]{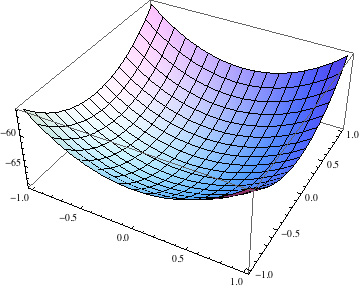}
(b)
\includegraphics[width=1.5in]{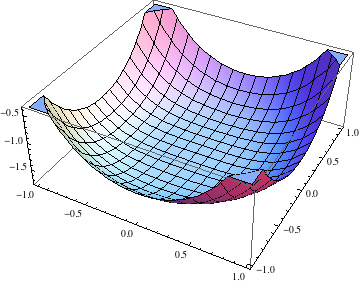}
(c)
\includegraphics[width=1.5in]{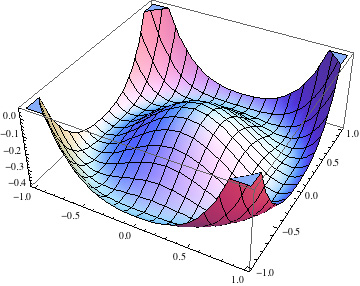}
\caption{Potential $V(\Phi)$ as function of $\Phi$. 
(a) $T=5$; (b) $T=2$; and (c) $T=1$.}\label{fig_potential} The plot shows the potential $V$ on the $z$-axis in term of the real and imaginary part of the scalar field $\Phi$
\end{figure}
In figure \ref{fig_potential} we show the potential as function of the complex scalar field (where we consider $m=1$ and $\lambda =1$). 
Units are such that $\hbar = 1$, $c = 1$, and $k_B = 1$.

The dynamics of a SF is governed by the Klein-Gordon (KG) equation, which comes by looking for the extremum of the action corresponding to the Lagrangian (\ref{eq:L}), it is the equation of motion of a field composed of spinless particles,
 \begin{equation}
 \Box_E^{2}\Phi+\frac{\partial V}{\partial \Phi^*}-2 m^2\phi{\Phi}=0,
 \label{eq:KG}
 \end{equation}
 where we have added a first order self-interaction potential  $\phi$ to the SF.
For a charged field the D'Alambertian operator is given by,
 \begin{equation}
  \Box_E
 ^{2}\equiv\left(\nabla_{\mu}+\mathrm{i} eA_\mu\right)\left(\nabla^{\mu}+\mathrm{i}eA^\mu\right)
  \label{eq:Box}
 \end{equation}
where 
$A_\mu=(\mathbf{A},\varphi)$
 is the electromagnetic four potential.
Observe that we can rewrite the D'Alambertian as
 \begin{equation}
  \Box_E^2=(\mathbf{\nabla}+
  \mathrm{i} e\mathbf{A})\cdot\mathbf{\nabla}-(\frac{\partial}{\partial t}+ 
  \mathrm{i} e\varphi)\frac{\partial}{\partial t}+\mathrm{i} e\nabla_\mu A^\mu- e^2A_\mu A^\mu
  \label{eq:Box}
 \end{equation}

In what follows we will use the Lorentz gauge $\nabla_\mu A^\mu=0$.
It is convenient to consider the total potential $V_T$ by adding to the potential $V$, the self-interaction contribution and the term $e^2A_\mu A^\mu=e^2A^2$, such that
  \begin{eqnarray}
   V_T(\Phi\Phi^*)&=& -m^2\Phi\Phi^*+\frac{\lambda}{4}T^2\Phi\Phi^*-e^2 A^2\Phi\Phi^*\nonumber\\&&+\frac{\lambda}{2}(\Phi\Phi^*)^2 -\frac{\pi^2}{90}T^4-2m^2\phi\, {\Phi\Phi^*} 
  \label{eq:VT}
  \end{eqnarray}
The KG equation now can be written as
 \begin{equation}
 \Box^2\Phi+\frac{\partial V_T}{\partial \Phi^*}=0,
 \label{eq:KGT}
 \end{equation}
 where now $\Box^2=\nabla_\mu\nabla^\mu$. 
Thus we can define an effective mass by
 \begin{equation}
 m_{eff}=\sqrt{m^2+e^2A^2}.
 \label{eq:meff}
 \end{equation}
 where the term coming from the electromagnetic potential term is the Proca mass.

Potential \eqref{eq:V} 
%
%
has a minimum at $\Phi=0$ when the temperature $T>T_c$. If $T<T_c$, the point $\Phi=0$ becomes a maximum and potential  \eqref{eq:V} has a circular zone of minima with radius
 \begin{eqnarray}
  \Phi_{min}
  &=&\frac{1}{2}\sqrt{T_c^2- T^2}
 \label{eq:min_phi}
 \end{eqnarray}
The critical temperature where the minimum of the potential $\Phi=0$ becomes a maximum and at which the symmetry is broken is (see figure \ref{fig_potential}), 
 \begin{equation}
  T_c=\frac{2}{\sqrt{\lambda}}\sqrt{m_{eff}^2+2 m^2\phi}.
 \label{eq:Tc}
 \end{equation}
This temperature is the point where the original $SU(1)$ symmetry of the Lagrangian is broken, because after this point the minimum of the potential is $\Phi_{min}$, where the Lagrangian does not contain  the $SU(1)$ symmetry anymore. See \cite{Elias2012} for details on BEC and more about this critical temperature.



\section{The generalized Gross-Pitaevskii equation}\label{sec:GP}

Now for the SF we perform the transformation 
 \begin{equation}
\Phi=\Psi\,\mathrm{e}^{-\mathrm{i} \hat m c t}\label{eq:Phi} 
  \nonumber ,
 \end{equation}
where $\hat{m}^2=m^2c^2/\hbar^2$. We have returned to normal units, except that temperature is in energy units ($k_B=1$).

In terms of function $\Psi$, the KG equation (\ref{eq:KG}) now reads,
\begin{eqnarray}
 \mathrm{i\hbar}\dot{\Psi}+\frac{\hbar^2}{2m}\Box^2{\Psi}&+&
\frac{\lambda}{2mc^2}|\Psi|^2\Psi-mc^2\phi\Psi+ec\varphi\Psi\nonumber\\
 &+&\frac{\lambda T^2}{8mc^2}\Psi=0,
\label{eq:GP}
\end{eqnarray}
the complex conjugate equation can be described in the same way. 
The notation used is: $\dot{\Psi}=\partial \Psi/\partial t$.

Equation
(\ref{eq:GP}) is the KG equation \eqref{eq:KG} or \eqref{eq:KGT} rewritten in terms of the function $\Psi$ and temperature $T$. 
This equation is an exact equation defining the field $\Psi(\mathbf{x},t)$, where $\phi$ is the external potential acting on the system and the terms in $\lambda$ represent the interaction potential within the system. 
When $T\rightarrow 0$ and in the non-relativistic limit, that means, in the limit where we take $c\gg 1$, the D' Alambertian $\Box^2\rightarrow\nabla^2$, eq. (\ref{eq:GP}) becomes the  Gross-Pitaevskii equation for Bose-Einstein Condensates (BEC) (see for example \cite{Guzman:2004wj}).
And also the static limit of equation (\ref{eq:GP}) is the known Ginzburg-Landau equation. 
Therefore we will consider equation (\ref{eq:GP}) as a generalization of the Gross-Pitaevskii equation 
that describe a complex charged scalar field in a thermal bath at finite temperature\cite{Elias2012}.



\section{The Hydrodynamical version}\label{sec:hydro}

In what follows we transform the generalized Gross-Pitaevskii equation (\ref{eq:GP}) into its analogous hydrodynamical version, 
for this purpose the ensemble wave function $\Psi$ will be represented in terms of a modulus $n$ and a phase $S$ as,
 \begin{equation}
 \Psi=\sqrt{ n}\,\mathrm{e}^{\mathrm{i}S}.
 \label{eq:psi}
 \end{equation}
where the phase $S(\mathbf{x},t)$ is taken as a real function. As usual this phase will define the velocity. 
We can interpret $n(\mathbf{x},t)=\rho/M_T$ as the ratio between the number density of particles in the condensed state, $\rho=mn_0=mN_0/L^3$, being $N_0$ the number of particles in condensed state and $M_T $ the total mass of the particles in the system. 
Both, $S$ and $ n$, are functions of time and position (See \cite{Almada} for a more detailed discussion on 
this hydrodynamical approach). 
The concept of SB is often used as a sufficient condition for BEC \cite{Elias2012}. 

So, as can be seen in figure \ref{fig_potential},
we have that the SF can oscillate around
the $\Phi=0$ minimum.
Below the critical temperature $T_c$, SF oscillates close to the ``ring'' minimum zone (For a real scalar field we will have that the oscillation ocurrs around the minimum values $\Phi_{min}^2=(T_c^2-T^2)/4$, and the density will oscillate around $n=\kappa^2 (T_c^2-T^2)/4$ as can be seen by equation (\ref{eq:min_phi})), where $\kappa$ is a scale constant, that will determined in some experiment.

Then, to obtain the hydrodynamical equations, we perform the Madelung transformation \eqref{eq:psi} in the generalized Gross-Pitaevskii equation \eqref{eq:GP}. We obtain:
\begin{subequations}
 \begin{eqnarray}
  \dot{ n}+\mathbf{\nabla}\cdot( n\mathbf{v})
  -\frac{\hbar}{2me}\dot j&=&0,\label{eq:cont}\\
  \dot{\mathbf{v}}+(\mathbf{v}\cdot\mathbf{\nabla})\mathbf{v}-\frac{ce}{m}\left(\mathbf{E}+\mathbf{v}\times\mathbf{B}\right)&=&\nonumber\\
  -c^2\mathbf{\nabla}\phi
  +\frac{\lambda}{m^2c^2\kappa^2}\mathbf{\nabla} n
  &+&\frac{\hbar^2}{m^2}\mathbf{\nabla}\left(\frac{\nabla^2\sqrt{ n}}{\sqrt{ n}}\right)\nonumber\\
  +\frac{\hbar}{2emc^2n}\dot{\mathbf{v}j}
  -\frac{\hbar^2}{m^2}\mathbf{\nabla}\left(\frac{\partial^2_t\sqrt{ n}}{\sqrt{ n}}\right)
  &+&\frac{\lambda}{4m^2}T\mathbf{\nabla}T\label{eq:2}
 \end{eqnarray}\label{eq:hydro}
\end{subequations}
where $\mathbf{E}$  
and $\mathbf{B}=\nabla\times\mathbf{A}$  are the electric and the magnetic field vectors, respectively. Notice that if we want to compare these equations with the hydrodynamical ones, there are extra terms corresponding with the relativistic character of the Klein-Gordon equation.  
We have also defined the fluxes
\begin{subequations}
\begin{eqnarray}
\mathbf{ j}=2e n(\nabla S+\frac{e}{\hbar}\mathbf{A})\label{eq:fluxesjvec}  , 
\quad 
j=2en(\dot S+\frac{e}{\hbar}\varphi)\label{eq:fluxesj} , 
\quad
j_\mu=(\mathbf{j},j)
\label{eq:fluxesjmu}
\end{eqnarray}
\label{eq:fluxes}
\end{subequations}
and
 \begin{equation}
  \mathbf{v}\equiv\frac{\hbar}{m}\left(\mathbf{\nabla}S+e\mathbf{A}\right)
 \label{eq:vel}
 \end{equation}

Notice that in (\ref{eq:2}) $\hbar$ enters on the right-hand side through the term containing the gradient of $ n$. 
This term is usually called the ``quantum pressure'' and is a direct consequence of the Heisenberg uncertainty principle, it reveals the importance of quantum effects in interacting gases.

Multiplying by $ n$, (\ref{eq:2}) can be written as:
 \begin{eqnarray}
  n\dot{\mathbf{v}}+n(\mathbf{v}\cdot\mathbf{\nabla})\mathbf{v}=n\mathbf{F}_E+n\mathbf{F}_\phi -\mathbf{\nabla}p
  +n\mathbf{F}_Q+\mathbf{\nabla}\sigma, 
 \label{eq:navier}
 \end{eqnarray}
where $\mathbf{F}_E=\frac{e}{m}\left(\mathbf{E}+\mathbf{v}\times\mathbf{B}\right)$ is the electromagnetic force, $\mathbf{F}_\phi=-\mathbf{\nabla}\phi$ is the force associated to the external potential $\phi$, $p$ can be seen as the pressure of the SF gas that satisfies the equation of state $p=wn^2$, $\mathbf{\nabla}p$ are forces due to the gradients of pressure and $w=-\lambda/4m^2$ is an interaction parameter, $\mathbf{F}_Q=-\mathbf\nabla U_Q$ is the quantum force associated to the quantum potential, 
\begin{equation}
 U_Q=-\frac{\hbar^2}{2m^2}\left(\frac{\nabla^2\sqrt{ n}}{\sqrt{ n}}\right),
 \label{eq:UQ}
 \end{equation}
and $\mathbf{\nabla}\sigma$ is defined as
\begin{eqnarray}
 \mathbf{\nabla}\sigma=\frac{\hbar}{2me}\dot{\mathbf{v}}j 
 &+&\frac{1}{4}\frac{\lambda}{m^2} nT\mathbf{\nabla}T\nonumber\\
 &+&\zeta\mathbf{\nabla}(\ln n{\dot)}
 -\frac{\hbar^2n}{4m^2}\mathbf{\nabla}\left(\frac{\ddot{n}}{n}\right),
 \label{eq:sigma}
 \end{eqnarray}
where the coefficient $\zeta$ is given by
\begin{equation}
 \zeta=\frac{\hbar^2}{4m^2}\left[-\mathbf\nabla\cdot( n\mathbf{v})+\frac{\hbar}{2me}j\right]\nonumber,
 \label{eq:zeta}
 \end{equation}
and the term $\nabla(\ln n{\dot)}$ can be written as
\begin{eqnarray}
 \mathbf{\nabla}(\ln n{\dot)}=-\mathbf{\nabla}(\mathbf{\nabla}\cdot\mathbf{v})
 -\mathbf{\nabla}[\mathbf\nabla(\ln n)\cdot\mathbf{v}]
 +\frac{1}{m}\mathbf{\nabla}[\frac{1}{ n}\dot j]\nonumber
 \label{eq:vrho1}
 \end{eqnarray}
System \eqref{eq:hydro} is the hydrodynamical representation to equation (\ref{eq:GP}) and is completely equivalent to it.



\section{The Newtonian limit}\label{sec:Newton}

  The Newtonian limit is characterized because we expand the time derivatives in order of a small parameter $1/\epsilon^2$, while the space derivatives are expanded in the same parameter as $1/\epsilon$. Neglecting second order time derivatives (terms with $1/\epsilon^4$ and beyond) and products of time derivatives we can simplify system (\ref{eq:hydro}) (observe that the velocity $\mathbf{v}$ and $\hbar$ are still second order). 
In  this limit we arrive to the non-relativistic system of equations \eqref{eq:hydro}, 
\begin{subequations}
 \begin{eqnarray}
  \dot{ n}+\mathbf{\nabla}\cdot( n\mathbf{v})&=&0,\label{eq:contNR}\\ 
  n\dot{\mathbf{v}}+n(\mathbf{v}\cdot\mathbf{\nabla})\mathbf{v}&=&n\mathbf{F}_E+n\mathbf{F}_\phi 
  -\mathbf{\nabla}p+n\mathbf{F}_Q+\mathbf{\nabla}\sigma.\nonumber\\
   \label{eq:navierNR}
 \end{eqnarray}\label{eq:hydroNR}
 \end{subequations}

Equation (\ref{eq:contNR}) is the continuity equation, and (\ref{eq:navierNR}) is the equation for the momentum transfer. Observe that this last one contains forces due to the external potential, to the gradient of the pressure, viscous forces due to the interactions of the condensate and forces due to the quantum nature of the equations. Quantity $\mathbf{\nabla}(\ln n{\dot)}$ plays a very important roll, in this limit it reads
\begin{equation}
 \mathbf{\nabla}(\ln n{\dot)}=-\mathbf{\nabla}(\mathbf{\nabla}\cdot\mathbf{v})
 -\mathbf{\nabla}[\mathbf{\nabla}(\ln n)\cdot\mathbf{v}]\nonumber.
 \label{eq:vrho1NR}
 \end{equation}
 Thus
\begin{eqnarray}
 \mathbf{\nabla}\sigma=\frac{1}{4}\frac{\lambda}{m} nT\mathbf{\nabla}T
 -\zeta\left[\mathbf{\nabla}(\mathbf{\nabla}\cdot\mathbf{v})
 +\mathbf{\nabla}[\mathbf{\nabla}(\ln n)\cdot\mathbf{v}]\right],
\label{eq:sigma2}
 \end{eqnarray}
where now we have
\begin{equation}
 \zeta=-\frac{\hbar^2}{4m^2}\mathbf{\nabla}\cdot( n\mathbf{v})\nonumber,
 \end{equation}

We interpret the function $\mathbf{\nabla}\sigma$ as the viscosity of the system, it contains terms which are gradients of the temperature and of the divergence of the velocity and density (dissipative contributions). The measurement of the temperature dependence in this thermodynamical quantity at the phase transitions might reveal important information about the behavior of the gas due to particle interaction.



\section{The Thermodynamics}\label{sec:thermodynamic}

In what follows we will derive the thermodynamical equations from the hydrodynamical representation. We can derive a conservation equation for a function $\alpha$, starting with the relationship \cite{Matos2011}
 \begin{equation}
  ( n\alpha)\dot{}= n\dot{\alpha}+\alpha\dot{n}
 \label{eq:phi1}
 \end{equation}
where $\alpha$ can take the values of $\phi$ and $U_Q$, both of them fulfil equation (\ref{eq:phi1}). Using the continuity equation 
  \begin{equation}
\dot{ n}+\mathbf{\nabla}\cdot( n\mathbf{v})=0
  \end{equation}
 in \eqref{eq:phi1} we obtain,
  \begin{equation}
   ( n\alpha)\dot{}+\mathbf{\nabla}\cdot(n\mathbf{v}\alpha)=-n\mathbf{v}\cdot\mathbf{F}_{\alpha}
   +n\dot{\alpha}\nonumber. 
   \label{eq:contphi}
  \end{equation}
where $\mathbf{F}_\alpha=-\nabla\alpha$. The treatment of 
$\sigma$ using the above this procedure is more difficult because in general we do not know it explicitly. 

 As we know, in general (for non-relativistic systems), the total energy density of the system $\epsilon$ is the sum of the kinetic, 
potential and internal energies 
in this case we have an extra term $U_Q$ due to the quantum potential,
 \begin{equation}
  \epsilon= ne=\frac{1}{2} nv^2+ n\phi+ nu+ nU_Q+\psi_E
 \label{eq:energiae}
 \end{equation}
being $u$ the internal energy of the system and 
 \begin{equation}
  \psi_E=\frac{e}{m}(\varphi-\mathbf{v}\cdot\mathbf{A})
 \label{eq:psiE}
 \end{equation}
 the electromagnetic energy potential, defined in terms of the vector potential $\mathbf{A}$ and the electric potential $\varphi$.

 Then from (\ref{eq:energiae}) we have that $u$ will satisfy the equation
  \begin{equation}
   ( nu)\dot{}+\mathbf{\nabla}\cdot\mathbf{J}_u-\mathbf{\nabla}\cdot\mathbf{J}_{\rho}+ n\dot{\phi}=
   -p\mathbf{\nabla}\cdot\mathbf{v},
    \label{eq:conte}
  \end{equation}
being $\mathbf{J}_u$ the energy current,  given by a energy flux and a heat flux, $\mathbf{J}_q$,
  \begin{equation}
   \mathbf{J}_u=n u\mathbf{v}+\mathbf{J}_q+\mathbf{J}_B-p\mathbf{v},\nonumber
   \label{eq:Je}
  \end{equation} 
where $\mathbf{\nabla}\cdot\mathbf{J}_q=\mathbf{v}\cdot(\mathbf{\nabla}\sigma)$, and $\mathbf{\nabla}\cdot\mathbf{J}_B=\mathbf{v}\cdot(n\mathbf{j}_B)$, being $\mathbf{j}_B$ given by the continuity equation of the vector potential $\mathbf{A}$
\begin{equation}
 \frac{\partial \mathbf{A}}{\partial t}+(\mathbf{v}\cdot\mathbf{\nabla}) \mathbf{A}=-(\mathbf{A}\cdot\mathbf{\nabla})\mathbf{v}+\frac{m}{e}\mathbf{j}_B,
   \label{eq:contA}
  \end{equation}
expressions that as we can see is related in a direct way to the velocity and gradients of temperature in the condensate, and is the one that shows in an explicit way the temperature dependence of the thermodynamical equations. With these definitions at hand we have,
  \begin{equation}
   \left(n u\right)\dot{}+\mathbf{\nabla}\cdot(n\mathbf{v}u+\mathbf{J}_q+\mathbf{J}_B-p\mathbf{v}
   -\mathbf{J}_\rho)+ n\dot{\phi}=-p\mathbf{\nabla}\cdot\mathbf{v}.
   \label{eq:contu}
  \end{equation}
where $\mathbf{J}_\rho=n\mathbf{v}_\rho=\hbar^2/4m^2(\mathbf{\nabla}\ln n)^\cdot$. In order to find the thermodynamical quantities of the system in equilibrium (taking $p$ as constant on a volume $L$), we restrict the system to the regime where the auto-interacting potential is constant in time, with this conditions at hand for (\ref{eq:contu}) we have:
\begin{equation}
   \left(n u\right)\dot{}+\mathbf{\nabla}\cdot(n\mathbf{v}u+\mathbf{J}_q+\mathbf{J}_B-p\mathbf{v}-\mathbf{J}_\rho)= 
   -p\mathbf{\nabla}\cdot\mathbf{v}
   \label{interna}
  \end{equation}
As always the first term will represent the change in the internal energy of the system, $-p\mathbf{\nabla}\cdot\mathbf{v}$ is the work done by the pressure and $\mathbf{\nabla}\cdot\mathbf{v}$ is related to the change in the volume.
The term with $\mathbf{J}_q$ contains terms related to the heat generated by gradients of the temperature $\mathbf{\nabla}T$ and dissipative forces due to viscous forces $\sim\mathbf{\nabla}(\mathbf{\nabla}\cdot\mathbf{v})$.
And finally but most important we have an extra term, $\mathbf{\nabla}\cdot\mathbf{J}_\rho$, due to gradients of the quantum potential (\ref{eq:UQ}).

 Integrating this 
 expression on a close region, we obtain
  \begin{eqnarray}
  \frac{\mathrm{d}}{\mathrm{d}t}\int n u\,\mathrm{d}V+\oint (\mathbf{J}_q+\mathbf{J}_B+p\mathbf{v})\cdot\mathbf{n}\, \mathrm{d}S
   &-&\oint\,\mathbf{J}_\rho\cdot\mathbf{n}\, \mathrm{d}S\nonumber\\ 
   &=&-p\frac{\mathrm{d}}{\mathrm{d}t}\int \,\mathrm{d}V\nonumber.
   \label{eq:contu2}
  \end{eqnarray}
Equation \eqref{interna} is the continuity equation for the internal energy of the system and as usual, from here we have an expression that would describe the thermodynamics of the system in an analogous way as does the first law of thermodynamics, in this case for the KG equation or a BEC. 
This reads
  \begin{equation}
   \mathrm{d}U=\text{\^d} Q+\text{\^d} Q_B+\text{\^d}A_Q-p\mathrm{d}V
   \label{eq:1leyBEC}
  \end{equation}
where $U=\int n u\,\mathrm{d}V$ is the internal energy of the system, 
and as we can see, its change is due to a combination of heat $Q$ added to the system and work done on the system (pressure dependent), and
  \begin{equation}
   \frac{\text{\^d}A_Q}{\mathrm{d}t}=\frac{\hbar^2}{4m^2}\oint\,n(\mathbf{\nabla}\ln n)\dot{}\cdot\mathbf{n}\,
   \ \mathrm{d}S=\oint n\mathbf{v}_\rho\cdot\mathbf{n}\,\mathrm{d}S,\nonumber
   \label{eq:AQ}
  \end{equation}
is the corresponding quantum heat flux due to the quantum nature of the KG equation. 

The first and third terms on the right hand side of equation \eqref{eq:1leyBEC} would make the crucial difference between a classical and a quantum first law of thermodynamics.
Analogously, for the magnetic heat we have
\begin{eqnarray}
   \frac{\text{\^d}Q_B}{\mathrm{d}t}&=&\int \mathbf{\nabla}\cdot\mathbf{J}_B\,dV= \int\mathbf{v}\cdot(n\mathbf{j}_B)\,dV\nonumber\\
  &=&\frac{m}{e}\int  n\left(\frac{\partial \mathbf{A}}{\partial t}+(\mathbf{v}\cdot\mathbf{\nabla}) \mathbf{A}+(\mathbf{A}\cdot\mathbf{\nabla})\mathbf{v}\right)\cdot\mathbf{v}\,dV\nonumber\\
   \label{eq:dQ_B}
  \end{eqnarray}
where the vector potential $\mathbf{A}$ fulfills the Maxwell equations, in terms of the fluxes \eqref{eq:fluxes} it reads
\begin{equation}
  {F^{\mu\nu}}_{,\nu}=- 
  j_\nu
   \label{eq:Maxwell}
  \end{equation}
where as usual $F_{\mu\nu}=\partial_\mu A_\nu-\partial_\nu A_\mu$.

To complete the description we write the Maxwell equations.
In terms of the vector and the electric potential, the Maxwell equations are given by
\begin{subequations}
\begin{eqnarray}
  {\Box}\mathbf{A}=-\frac{1}{\kappa^2}\mathbf{j}\label{eq:Maxwellvec}\\
   {\Box}\varphi=-\frac{1}{\kappa^2}j
   \label{eq:Maxwellesc}
  \end{eqnarray}\label{eq:Maxwell2}
  \end{subequations}
where we have used the Lorentz gauge. Observe that the fluxes contain the information of the velocity of the fluid and of the electromagnetic term as well.


\section{Conclusions}\label{sec:Conclusions}
In this work we have studied the $U(1)$ SB of the Klein-Gordon Lagrangian with temperature contributions to the effective Mexican hat potential of a system of weakly interacting bosons. We did this using a simple transformation in order to rewrite the Klein-Gordon equation as a GP-like one. This is not surprising, because the Klein-Gordon equation represents bosons with spin zero and the GP equation represents bosons in condensed state. We interpret this GP-like equation as a generalization of the normal GP one but for relativistic particles and finite temperature. 
Using the Madelung transformation we rewrite again the Klein-Gordon equations in their hydrodynamical version where we have obtained several thermodynamical relations for the Gross-Pitaevskii equation (\ref{eq:GP}). With this thermodynamical version it is much simpler to interpret the physics of the phenomena happening during the SB. For example we find that in certain conditions the system becomes super-conductor and under other conditions it becomes super-fluid. In this interpretations super-conductivity and super-fluidity can be understood from first principles of quantum field theory in a unified way.



\end{document}